\documentstyle[12pt]{article}
\title{On a gauge invariant quantum formulation for non-gauge
classical theory}
\author{I.L. Buchbinder\\
\it Department of Theoretical Physics, \\
\it Tomsk State Pedagogical University, \\
\it Tomsk 634041, Russia
\and
V.D. Pershin\\
\it Department of Theoretical Physics, \\
\it Tomsk State University,\\
\it Tomsk 634050, Russia
\and
G.B. Toder\\
\it Department of Physics and Chemistry, \\
\it Omsk State Academy of Railway Communications,\\
\it Omsk 644046, Russia}
\date{}
\begin{document}
\begin{titlepage}
\maketitle
\thispagestyle{empty}

\begin{abstract}
We propose a method of constructing a gauge invariant canonical
formulation for non-gauge classical theory which depends on a
set of parameters. Requirement of closure for algebra of
operators generating quantum gauge transformations leads to
restrictions on parameters of the theory. This approach is then
applied for illustration to bosonic string theory coupled to
background tachyonic field. It is shown that within the proposed
canonical formulation the known mass-shell condition for tachyon
is produced.
\end{abstract}
\end{titlepage}

{\bf 1.} The procedure of canonical quantization provides a
natural and consistent approach to consruction of quantum models
in modern theoretical physics. BFV method \cite{BFV} is the most
general realization of this procedure ensuring unitarity at the
quantum level and consistency of theory symmetries and dynamics.
Now BFV method has been studied in details \cite{BatFr,Hen} yet
formulation of new models in quantum field theory requires
investigation of unexplored aspects of canonical formulation.

In this paper we discuss one of these aspects arising from
bosonic string theory coupled to background fields \cite{fri}. A
crucial point of string models is requirement of conformal
invariance at the quantum level. It leads to restrictions on
spacetime dimension in the case of free string theory
\cite{Pol,Hwang} and to effective equations of motion for massless
background fields in the case of string theories coupled to
background \cite{Lov}. In terms of covariant functional methods
this condition appears as independence of quantum effective
action on the conformal factor of two-dimensional metrics or as
vanishing of renormalized operator of the energy-momentum trace.

According to the prescription \cite{Pol} generally accepted in
functional approaches to string theory dynamical
variables should be treated in different ways. Namely,
functional integration is carried out only over string
coordinates $X^\mu(\tau,\sigma)$ while components of
two-dimensional metrics $g_{ab}(\tau,\sigma)$ are considered as
external fields. Then one demands the result of such an
integration to be independent on the conformal factor and the
integrand over $g_{ab}(\tau,\sigma)$ reduces to finite
dimensional integration over parameters specifying string world
sheet topologies. This prescription differs from the standard
field theoty rules when functional integral is calculated over
every variable independently \footnote{In string theory this
independent integration would lead to appearence at the quantum
level of an extra degree of freedom connected with
two-dimensional gravity \cite{buchshap}.}.

This approach can be as well applied to string theory
interacting with massive background fields which is not
classically conformal invariant. One demands that operator of
the energy-momentum vanish no matter whether the corresponding
classical action is conformal invariant or not. As was shown in
\cite{buchper} it gives rise to effective equations of motion
for massive background fields. It means that a non-gauge
classical theory depending on a set of parameters is used for
constructing of a quantum theory that is gauge invariant under
some special values of the parameters. Such a situation occurs
in string theory if interaction with massless dilaton, tachyon
or any other massive field is turned on.

The problem is how to describe this procedure in terms of
canonical quantization. Due to the general BFV method one should
construct hamiltonian formulation of classical theory,
find out all constraints and calculate algebra of their Poisson
brackets. Then one defines fermionic functional $\Omega$
generating algebra of gauge transformations and bosonic
functional $H$ containing information of theory dynamics.
Quantum theory is consistent provided the operator $\hat\Omega$
is nilpotent and conserves in time. The corresponding analysis
for bosonic string coupled to massless background fields was
carried out in  \cite{buch91,buch95}.

In the case of string theory interacting with massive background
fields components of two-dimensional metrics should be treated
as external fields, otherwise classical equations of
motion would be inconsistent. As a consequence, classical
gauge symmetries are absent and it is impossible to construct
classical gauge functional $\Omega$. In this paper we propose a
prescription allowing for some models to construct quantum
operator $\hat\Omega$ starting with a classical theory without
first class constraints. Quantum theory is gauge invariant if
there exist values of theory parameters providing nilpotency and
conservation of operator $\hat\Omega$. Then to illustrate how
the prescription works we apply it to the theory of closed
bosonic string coupled to tachyonic field.

\medskip

{\bf 2.} Consider a system described by a hamiltonian
\begin{equation}
H=H_{0}(a)+\lambda^{\alpha}T_{\alpha}(a)
\label{Hgen}
\end{equation}
where $H_0(a)=H_0(q,p,a)$, $T_{\alpha}(a)=T_{\alpha}(q,p,a)$ and
$q$, $p$ are canonically conjugated dynamical variables;
$a=a_i$ and $\lambda^{\alpha}$ are external parameters of the
theory.

We suppose that $T_{\alpha}(a)$ are some functions of the form
\begin{equation}
T_{\alpha}(a)=T^{(0)}_{\alpha}(a)+T^{(1)}_{\alpha}(a)
\end{equation}
and closed algebra in terms of Poisson brackets is formed by
$T^{(0)}_{\alpha}(a)$, not by $T_{\alpha}(a)$:
\begin{eqnarray}
\{ T^{(0)}_{\alpha}(a),T^{(0)}_{\beta}(a) \} & = &
T^{(0)}_{\gamma}(a)U^{\gamma} _{\alpha\beta}(a)
\nonumber
\\
\{ H_0(a),T^{(0)}_{\alpha}(a) \} & = &
T^{(0)}_{\gamma}(a)V^{\gamma} _{\alpha}(a)
\end{eqnarray}
Such a situation may occur, for example, if
$T^{(0)}_{\alpha}(a)$ correspond to a free gauge invariant
theory and $T^{(1)}_{\alpha}(a)$ describe a perturbation
spoiling gauge invariance.

At the quantum level both the algebras of
$T^{(0)}_{\alpha}(a)$ and $T_{\alpha}(a)$ are not closed in
general case
\begin{eqnarray}
[T_{\alpha}(a), T_{\beta}(a)] &=& i\hbar \bigl(
T_{\gamma}(a) U^{\gamma} _{\alpha\beta}(a) + A_{\alpha\beta}(a)
\bigr),
\nonumber
\\
{} [H_0(a),T_{\alpha}(a)]& = & i\hbar \bigl(
T_{\gamma}(a)V^{\gamma}_{\alpha}(a) + A_{\alpha}(a) \bigr),
\label{A}
\end{eqnarray}
and operators $A_{\alpha\beta}$, $A_\alpha$ do not vanish in the
limit $\hbar\rightarrow 0$ due to absence of classical gauge
invariance.

We define quantum operators $\Omega$ and  $H$ as follows:
\begin{eqnarray}
\Omega & = &
c^{\alpha}T_{\alpha}(a)-{1\over2}U^{\gamma}_{\alpha\beta}(a)
:{\cal P_{\gamma}} c^{\alpha} c^{\beta}:
\nonumber
\\
H & = & H_0 (a) + V^{\gamma}_{\alpha}(a) :{\cal P_{\gamma}}
c^{\alpha}:
\end{eqnarray}
where $:\quad:$ stands for some ordering of ghost fields. Square
of such an operator $\Omega$ and its time derivative take the
form
\begin{eqnarray}
\Omega^{2} & = & {1\over2}i\hbar
\bigl( A_{\alpha\beta}(a) + G_{\alpha\beta}(a) \bigr)
:c^{\alpha}c^{\beta}:
\nonumber \\
\frac{d\Omega}{dt}& = &
\frac{\partial\Omega}{\partial t}+[H,\Omega] = {}
\nonumber \\
& = & \left( \frac{\partial T_{\alpha}(a)}{\partial t}-
A_{\alpha}(a) - G_{\alpha}(a)\right) c^{\alpha}-{1\over2}
\frac{\partial U^{\gamma}_{\alpha\beta}(a)}{\partial t}
:{\cal P_{\gamma}}c^{\alpha}c^{\beta}:
\label{Omega}
\end{eqnarray}
where $G_{\alpha\beta}(a)$, $G_\alpha(a)$ are possible quantum
contributions of ghosts.

It is natural to suppose that every operator of the theory can
be performed as a linear combination of an irreducible set of
independent operators $O_M(q,p)$:
\begin{eqnarray}
& & A_{\alpha\beta}(a)+G_{\alpha\beta}(a) = E^{M}_{\alpha\beta}(a)O_{M}(q,p)
\nonumber \\
& &\frac{\partial T_{\alpha}(a)}{\partial t} - A_{\alpha}(a) - G_{\alpha}(a)
= E^{M}_{\alpha}(a)O_{M}(q,p)
\end{eqnarray}
$E^{M}_{\alpha\beta}(a)$, $E^M_\alpha(a)$ being some $c-$valued
functions of the parameters $a$.

In general case $\Omega^2 \neq 0$ and $d\Omega/dt \neq 0$.
However, if equations
\begin{equation}
E^{M}_{\alpha\beta}(a) = 0, \quad
E^{M}_{\alpha}(a) = 0, \quad
\frac{\partial U^{\gamma}_{\alpha\beta}(a)}{\partial t}=0,
\label{a_0}
\end{equation}
have some solutions $a_i=a_i^{(0)}$ then operator $\Omega$ is
nilpotent and conserves for these specific values of parameters
and hence the quantum theory is gauge invariant. Thus, there
exists a possibility to construct quantum theory with given gauge
invariance that is absent at the classical level.

It is important that eqs.(\ref{a_0}) are not anomaly
cancellation conditions because an anomaly represents breaking
of classical symmetries at the quantum level. In the theory
under consideration classical symmetries are absent and
eqs.(\ref{a_0}) express conditions of quantum symmetries
existence.

In specific models eqs.(\ref{a_0}) may have no solutions at all
or, conversely, be fulfilled identically. The latter possibility
was described in \cite{Fubini}.

\medskip

{\bf 3.} As  an example where the described procedure really
works and leads to eqs.(\ref{a_0}) with non-trivial solutions
for parameters $a$ we consider closed bosonic string theory
coupled with massive tachyonic fields. Structure of this theory
has been studied in details within covariant functional methods
and we are going to show that our procedure reproduces the
correct equation for tachyonic field.

The theory is described by the classical action
\begin{equation}
S=-{1\over 2\pi\alpha'} \int d^{2}\sigma\,
  \sqrt{-g} \bigl\{{1\over 2} g^{ab}\partial_{a} X^{\mu} \partial_{b}
  X^{\nu} \eta_{\mu\nu} +Q(X)\bigr\},
\end{equation}
$\sigma^a=(\tau,\sigma)$ are coordinates on string world sheet,
$\eta_{\mu\nu}$ is Minkowski metrics of $D-$dimensional
spacetime and $Q(X)$ is background tachyonic field.

If components of two-dimensional metrics $g_{ab}$ were
considered as independent dynamical variables the corresponding
classical equations of motion would be fulfilled only for
vanishing tachyonic fields:
\begin{equation}
g_{ab} \frac{\delta S}{\delta g_{ab}}=-
{1\over 2\pi\alpha'}\sqrt{-g}Q(X)=0.
\end{equation}
Similar situation arises for string theories coupled to another
massive fields \cite{buchper} or to massless dilaton \cite{buch91}.

Hence, to construct a theory with non-trivial tachyon we have to
treat components of $g_{ab}$ as external fields. This treatment
quite corresponds to covariant methods where functional integral
is calculated only over $X^\mu$ variables.

After the standard parametrization of metrics
\begin{eqnarray}
g_{ab}& = &e^{\gamma}\left(\begin{array}{cc}
\lambda^{2}_{1}-\lambda^{2}_{0} & \lambda_{1} \\
\lambda_{1}  & 1 \end{array}\right)
\end{eqnarray}
the hamiltonian takes the form
\begin{equation}
H=\int d\sigma\,(p_\mu {\dot x}^\mu - L)
=\int d\sigma \,(\lambda_{0}T_{0}+\lambda_{1}T_{1}),
\label{H}
\end{equation}
where
\begin{eqnarray}
T_0 & = & T_0^{(0)} + {1\over 2\pi\alpha'}e^{\gamma}Q(X), \quad
T_0^{(0)} ={1\over2}\left( 2\pi\alpha' P^2 +
{1\over 2\pi\alpha'} X'^2 \right),
\nonumber \\
T_1 & = & T_1^{(0)} = P_{\mu}X'^\mu {,}
\label{T}
\end{eqnarray}
$P_\mu$ are momenta canonically conjugated to $X^\mu$,
$X'^\mu=\partial X^\mu/\partial\sigma$; $T_0^{(0)}$
and $T_1^{(0)}$ represent constraints of free string theory and
form closed algebra in terms of Poisson brackets. $\lambda_0$
and $\lambda_1$ play the role of external fields and so $T_0$
and  $T_1$ cannot be considered as constraints of classical
theory. In free string theory conditions $T^{(0)}_0=0$,
$T^{(0)}_1=0$ result
from conservation of canonical momenta conjugated to
$\lambda_0$ and $\lambda_1$. According to our prescription in
string theory with tachyon $\lambda_0$ and $\lambda_1$ can not
be considered as dynamical variables, there are no corresponding
momenta and conditions of their conservation do not appear.

The role of parameters $a$ in the theory under consideration is
played by tachyonic field $Q(X)$ and conformal factor of
two-dimensional metrics $\gamma(\tau,\sigma)$. The theory (\ref{H})
is of the type (\ref{Hgen}) with $H_0=0$, structural
constants of classical algebra being independent on time. It
means that the condition of conservation for the operator
$\Omega$ (\ref{Omega}) will be met if operators (\ref{T}) do not
depend on time explicitly, that is
\begin{equation}
\dot{\gamma}(\tau,\sigma) = {\partial\gamma(\tau,\sigma)\over\partial\tau}=0.
\end{equation}

To derive the first condition (\ref{a_0}) one has to construct
quantum algebra of operators $T_0$, $T_1$. We will use Fourier
components of the operators
\begin{eqnarray}
L_n & = & \int^{2\pi}_0 d\sigma\, e^{-in\sigma}
{1\over 2} (T_0-T_1) ,
\nonumber \\
\bar{L}_n & = & \int^{2\pi}_0 d\sigma\, e^{in\sigma}
{1\over 2} (T_0+T_1) .
\label{L_n}
\end{eqnarray}
Direct calculation of commutators of the operators (\ref{L_n})
gives the following algebra:
\begin{eqnarray}
[L_n , L_m ] & = & \hbar (n-m)L_{n+m} +
\hbar^2 \delta_{n,-m}  \left( {D\over 12}n(n^2 -1) + 2\alpha(0)n \right)
\nonumber \\
& &{}- (4\pi\alpha' )^{-1}\hbar (n-m)\int^{2\pi}_0
d\sigma\, e^{-i(n+m)\sigma} e^{\gamma (\tau , \sigma)}
\left( 1+\alpha'\hbar\partial^2  /4\right) Q(X),
\nonumber \\
{}[\bar{L}_n , \bar{L}_m ] & = & \hbar (n-m)\bar{L}_{n+m} +
\hbar^2 \delta_{n,-m}  \left( {D\over 12}n(n^2 -1) + 2\beta(0)n \right)
\nonumber \\
& & {}-(4\pi\alpha' )^{-1}\hbar (n-m)\int^{2\pi}_0
d\sigma\, e^{i(n+m)\sigma} e^{\gamma (\tau , \sigma)}
\left( 1+\alpha'\hbar\partial^2 /4\right) Q(X),
\nonumber \\
{}[L_n , \bar{L}_m ] & = & {}-(4\pi\alpha^\prime )^{-1}\hbar
(n-m)\int^{2\pi}_0  d\sigma\, e^{i(m-n)\sigma} e^{\gamma
(\tau , \sigma)} \left( 1+\alpha' \hbar\partial^2 /4\right)Q(X)
\nonumber \\
& &{}- (4\pi\alpha' )^{-1} i\hbar
 \int^{2\pi}_0 \, d\sigma\, e^{i(m-n)\sigma}
e^{\gamma (\tau,\sigma)} \gamma'(\tau,\sigma) Q(X){,}
\label{LL}
\end{eqnarray}
where $\alpha(0)$, $\beta(0)$ are constant parameters describing
ordering ambiguity. We see that the only quantum contributions to
the comutators are these of the second order in $\hbar$, all the
higher contributions equal zero.

Eqs.(\ref{LL}) define the explicit form of the function
$A_{\alpha\beta}$ (\ref{A}) in the string theory with tachyon.
Ghost contribution $G_{\alpha\beta}$ has the same structure as
in free string theory and cancels the $c-$valued terms in
(\ref{LL}) provided  that $D=26$ and
$\alpha(0)=\beta(0)=1$.

As a result the eqs.(\ref{a_0}) for string theory coupled to
tachyon appear as
\begin{equation}
D=26, \qquad \alpha(0)=1, \qquad \beta(0)=1,
\end{equation}
\begin{equation}
(\partial^2 + 4/\alpha'\hbar)Q(X)=0
\label{shell}
\end{equation}
\begin{equation}
\dot\gamma(\tau,\sigma)=0, \qquad \gamma'(\tau,\sigma)=0.
\label{gamma}
\end{equation}
The eqs.(\ref{shell}) is mass-shell condition for free tachyon.
It is linear and so effects of tachyonic intaraction are not
reproduced within perturbative calculation of the conformal
algebra (\ref{LL}). This is quite natural because taking into
account tachyonic interaction is known to require
non-perturbative approaches \cite{Das}.

The eqs.(\ref{gamma}) show that the operators $L_n$, $\bar{L}_n$
form conformal algebra only in the case $\gamma=const$. This
does not contradict the corresponding results of covariant
approaches. After integration over $X^\mu$ the effective action
does not depend on conformal factor provided effective equations
of motion for background fields are fulfilled. It means that any
physical results (e.g. values of correlation functions) do not
depend on a gauge choice of $\gamma$. Specifically, $\gamma$  can
be choosen to be constant and this situation is reproduced on
our case. It should be noted that our approach is not restricted
by flat world sheets. Quantum theory can be formulated for any
functions $\gamma$ but it is conformal invariant only for
constant $\gamma$.

The described example demonstrates a possibility to construct
canonical formulation of quantum theory invariant under gauge
transformations that are absent at the classical level. The
proposed method opens up a possibility for deriving interacting
effective equations of motion for massive and massless
background fields within the framework of canonical formulation
of string models and provides a justification of covariant
functional approach to string theory.

\medskip

The authors are grateful to I.V.~Tyutin and P.M.~Lavrov for
discussions of some aspects of the paper. The work was supported
by International Scientific Foundation, grant No RI~1300, and
Russian Foundation for Fundamental Research, project No
96-02-16017. The work was completed during visit of one of the
authors (ILB) to Institute of Physics, Humboldt Berlin
University supported by Deutsche Forschungsgemeinschaft,
contract DFG-43RUS113. ILB is grateful
to D.~L\"ust, D.~Ebert, H.~Dorn, G.~Cardoso, C.~Preitschopf,
C.~Schubert and M.~Schmidt for interesting discussions.


\newpage

\begin{thebibliography}{50}
\bibitem{BFV}
 E.S.Fradkin, G.A.Vilkovisky, Phys.Lett. {\bf 55B}, 224, (1975);
 I.A.Batalin, G.A.Vilkovisky, Phys.Lett. {\bf 69B}, 409, (1977);
 E.S.Fradkin, T.E.Fradkina, Phys.Lett. {\bf 72B}, 343, (1978).
\bibitem{BatFr}
 I.A.Batalin, E.S.Fradkin, Phys.Lett. {\bf 128B}, 303, (1983);
  J.Math.Phys {\bf 25}, 2426, (1984);
  Riv.Nuovo Cim. {\bf 9}, 1, (1986);
  Phys.Lett. {\bf 180B}, 157, (1986);
  Nucl.Phys. {\bf 179B}, 514, (1987);
  Ann.Inst.H.Poincare, {\bf 49}, 145, (1988).
\bibitem{Hen}
 M.Henneaux, Phys.Rept. {\bf 126}, 1, (1985);
 M.Henneaux, C.Teitelboim. Quantization of Gauge Systems.
  Princeton Univ.Press., Princeton, N.J., 1992.
\bibitem{fri}
  D.Friedan, Ann.Phys.(USA) {\bf 163}, 318, (1985);
  S.Mukhi, Nucl.Phys. {\bf B264}, 640, (1986).
 E.S.Fradkin, A.A.Tseytlin, Phys.Lett. {\bf 158B}, 316, (1985),
 Nucl.Phys. {\bf 261B}, 1, (1985).
\bibitem{Pol}
 A.M.Polyakov, Phys.Lett. {\bf B103}, 207, (1981).
\bibitem{Hwang}
 S.Hwang, Phys.Rev. {\bf D28}, 2614, (1983).
\bibitem{Lov}
 C.Lovelace, Phys.Lett. {\bf 135B}, 75, (1984);
 C.G.Callan, O.Friedan, E.Martinec, M.J.Perry,
    Nucl.Phys. {\bf 262B}, 593, (1985);
 A.A.Tseytlin, Nucl.Phys. {\bf 294B}, 383, (1987);
 H.Osborn, Nucl.Phys. {\bf 294B}, 595, (1987).
\bibitem{buchshap}
 I.L.Buchbinder, I.L.Shapiro, S.G.Sibiryakov,
   Nucl.Phys. {\bf 445B}, 109, (1995).
\bibitem{buchper}
 I.L.Buchbinder, E.S.Fradkin, S.L.Lyakhovich, V.D.Pershin,
   Phys.Lett. {\bf 304B}, 239, (1993);
 I.L.Buchbinder, V.A.Krykhtin, V.D.Pershin, Phys.Lett. {\bf 348B},
 63, (1995), Phys.Atom.Nucl. (Yad.Fiz.) {\bf 59}, 332 (1996).
\bibitem{buch91}
 I.L.Buchbinder, E.S.Fradkin, S.L.Lyakhovich, V.D.Pershin, Int.J.Mod.Phys.
 {\bf 6A}, 1211, (1991).
\bibitem{buch95}
 I.L.Buchbinder, B.R.Mistchuk, V.D.Pershin, Phys.Lett. {\bf 353B}, 257,
 (1995).
\bibitem{Fubini}
 S.Fubini, M.Roncadelli, Phys.Lett. {\bf 203B}, 433, (1988).
\bibitem{Das}
 S.R.Das, B.Sathiapalan, Phys.Rev.Lett. {\bf 56}, 2654, (1986);
 C.Itoi, Y.Watabiki, Phys.Lett. {\bf 198B}, 486, (1987);
 A.A.Tseytlin, Phys.Lett. {\bf 264B}, 311, (1991).
\end{thebibliography}
\end{document}